\documentclass{llncs}
\pdfoutput=1
\usepackage[utf8]{inputenc}
\usepackage{amssymb}
\usepackage{amsmath}
\usepackage{graphicx}
\usepackage{float}
\usepackage{url}

\begin{document}

\title{Data-Driven Relevance Judgments for Ranking Evaluation}

\author{Nuno Moniz\inst{1} \and Luís Torgo\inst{1} \and João Vinagre\inst{1}}
\institute{LIAAD - INESC Tec \\
Sciences College - University of Porto\\
Porto, Portugal\\
\email{nmmoniz@inescporto.pt, ltorgo@dcc.fc.up.pt, jnsilva@inesctec.pt}}

\maketitle
\begin{abstract}

Ranking evaluation metrics are a fundamental element of design and improvement efforts in information retrieval. We observe that most popular metrics disregard information portrayed in the scores used to derive rankings, when available. This may pose a numerical scaling problem, causing an under- or over-estimation of the evaluation depending on the degree of divergence between the scores of ranked items. The purpose of this work is to propose a principled way of quantifying multi-graded relevance judgments of items and enable a more accurate penalization of ordering errors in rankings. We propose a data-driven generation of relevance functions based on the degree of the divergence amongst a set of items' scores and its application in the evaluation metric Normalized Discounted Cumulative Gain ($nDCG$). We use synthetic data to demonstrate the interest of our proposal and a combination of data on news items from Google News and their respective popularity in Twitter to show its performance in comparison to the standard $nDCG$. Results show that our proposal is capable of providing a more fine-grained evaluation of rankings when compared to the standard $nDCG$, and that the latter frequently under- or over-estimates its evaluation scores in light of the divergence of items' scores.
\end{abstract}

\section{Introduction}

A crucial element of a ranking function is its evaluation design. Research shows that for different types of applications, no single optimal ranking metric is robust enough to work generally~\cite{Croft2009}, and that all metrics disagree to some extent when relating rankings and user preferences.

Consider the normalized form of the rank-based metric Discounted Cumulative Gain $DCG$ ($nDCG$)~\cite{Jarvelin2000}, considered to be robust~\cite{Sanderson2010}. It allows users to associate multi-graded relevance judgments to items presented in a ranking, deviating from the common binary notion of relevance. It also uses discount factors to simulate user experience by decreasing the impact of the evaluation of a given item as you go through the ranking in a descending fashion, while most metrics weight the positions uniformly. Despite the robustness of the metric, issues have been raised regarding its ad-hoc definition of relevance~\cite{Voorhees2001}.

In scenarios where baseline items' scores are available and the divergence of such scores is considered an important factor, the evaluation design may require a greater detail on the impact of ranking errors. Such requirements may be found in scenarios such as the recommendation of news based on their expected social popularity or the recommendation of stock options based on future winnings. Given the issues raised by the standard ad-hoc definition of relevance~\cite{Voorhees2001}, these scenarios may require a principled manner of determining multi-graded relevance judgments of items in a finer granularity.

Consider the example of a news recommender system focused on anticipating the most popular daily news and suggesting them to users. This scenario is depicted in Table~\ref{tbl:example}, where the popularity score predicted by the system and the final popularity score are graded between $1$ and $100$. In the example, the system recommends five news items on the first day and five other in the second day. 

\addtolength{\tabcolsep}{5pt}
\begin{table}[!h]
\centering
\caption{Illustrative example of the suggestions made by a news recommender system (RecSys) based on popularity scores.}\label{tbl:example}
\begin{tabular}{|c|c|c|c|c|c|c|c|c|}
\cline{1-4} \cline{6-9}
\multicolumn{4}{|c|}{Day 1} &  & \multicolumn{4}{c|}{Day 2} \\ \cline{1-4} \cline{6-9}
\# & News       & RecSys  & Baseline &  & \# & News       & RecSys & Baseline \\ \cline{1-4} \cline{6-9}
1 & $N_{8}$   & 70      & 100  & & 1 & $N_{102}$ & 70     & 100  \\
2 & $N_{5}$   & 50      & 80   & & 2 & $N_{114}$ & 50     & 20   \\
3 & $N_{2}$   & 40      & 90   & & 3 & $N_{107}$ & 40     & 90   \\
4 & $N_{11}$  & 15      & 15   & & 4 & $N_{129}$ & 15     & 15   \\
5 & $N_{10}$  & 10      & 10   & & 5 & $N_{139}$ & 10     & 10    \\ \cline{1-4} \cline{6-9}
\end{tabular}
\end{table}
\addtolength{\tabcolsep}{-5pt}

According to the final scores, the system made the same ranking error on both days by swapping the items on positions 2 and 3. Although the system made the same ordering error in both rankings, the error in the second day is considered more serious than the first. While permuting $N_5$ and $N_2$ seems acceptable given their similar final popularity score, the same does not happen when permuting $N_{114}$ and $N_{107}$. This motivates the issue of how to correctly estimate the impact of ordering errors in ranking evaluation for such scenarios. Since most ranking evaluation metrics disregard information portrayed in the scores (\textit{e.g.} distribution), they may over- or under-estimate rankings errors.

In this paper we show that the ad-hoc definition of relevance in $nDCG$ poses issues of under- and over-estimation in the calculation of the metric, due to the disregard of the numerical scale of the ranked items' scores. We then propose an automatic approach to attribute relevance judgments to items, accounting for the differences between their scores. Our approach is evaluated in a synthetic and real-world scenarios and the differences in evaluation impact between the ad-hoc definition of relevance and our approach are discussed.

For the remainder of this paper we will use the terminology described in Table~\ref{tbl:terminology}. The paper is structured as follows. In Section~\ref{sec:dcg} the $DCG$ metric is described. The proposal to tackle the numerical scaling issue is presented in Section~\ref{sec:solution}. An experimental evaluation concerning the application of our proposal in the evaluation metric $nDCG$ is described in Section~\ref{sec:exps}. A discussion is provided in Section~\ref{sec:discussion} and conclusions presented in Section~\ref{sec:conclusions}.

\begin{table}[!h]
\centering
\scriptsize
\caption{List of terms used in the paper.}
\label{tbl:terminology}
\begin{tabular}{|l|l|}
\hline
\textbf{Term}          & \textbf{Description}                                                                   \\ \hline
Item score ($y$)         & Numeric value associated to a given item ($x$), used to derive the ranking.                                              \\ \hline
Relevance judgment ($v$) & Property of an item depicting its level of relevance.                                  \\ \hline
Gain value ($g$)         & Parameter of $DCG$ given by a function of the relevance judgments.                      \\ \hline
Discount factor ($d$)    & Parameter of $DCG$ associated to the probability of a user seeing\\ & an item in a ranking. \\ \hline
\end{tabular}
\end{table}

\section{Discounted Cumulative Gain}\label{sec:dcg}

The Discounted Cumulative Gain (DCG)~\cite{Jarvelin2000} is composed of two sets of parameters: gain values and discount factors. The gain values are given by a function of ad-hoc relevance judgments associated to items in the ranking (\textit{i.e.} the higher the relevance, the higher the gain value, and vice-versa.). The discount factor is associated with the fact that the probability of a user seeing an item decreases as you go through the positions in a ranking. In essence, DCG is a weighted (by the discount factor $d$) sum of the level of relevancy (given by the gain values $g$) of the items in a rank $R$ of size $n$, i.e. $DCG_{d,g,n}(R) = \sum_{i=1}^{n} d_i g_i$.

This metric is commonly used in its normalized form, Normalized Discounted Cumulative Gain ($nDCG$). The normalization is carried out as follows:

\begin{equation}
nDCG(R) = \frac{DCG(R)}{IdealDCG(R)}
\end{equation}

\noindent where the $IdealDCG(R)$ represents the maximum possible $DCG(R)$  given by the optimal ranking order.

In literature concerning $DCG$ the two most popular forms of obtaining gain values ($g$) are: \textit{1)} by using $g_i = v_i$, and \textit{2)} through an exponential approach $g_i = 2^{v_i} - 1$, where $v_i$ is the relevance judgment associated with a given item of index $i$. 
To avoid ad-hoc relevance settings, proposals have been made for data-driven approaches for learning gain values. Zhou et al.~\cite{Zhou2014} propose a methodology capable of learning the gain values and discount factors from paired preferences of users. The authors modeled the problem as a special case of learning linear utility functions. Demeester et al.~\cite{Demeester2016} propose the predicted relevance model (PRM) to capture differences between assessor judgments and estimate the relevance of documents for random users. Although related, these approaches are focused on learning the relevance of items by paired preferences of users. In our work we focus on problems where baseline scores are available (\textit{e.g.} retrieval scores) and the problem is correctly determining the relevance judgment of items. We will base our work on the most common form of obtaining gain values: the exponential approach.

Regarding the discount factor ($d$), according to Wang et al.~\cite{Wang2013}, there are at least two discount factors that guarantee the notion of consistent distinguishability. This notion states that for every pair of substantially different ranking functions, the ranking measure can decide which one is better in a consistent manner on almost all datasets. The authors concluded that both the logarithmic discount $\frac{1}{log(i+1)}$ and the Zipfian discount $i^{-1}$~\cite{Kanoulas2009}, where $i$ is the position of an item in the ranking, guarantee this property. In this paper we will assume the most commonly used, the logarithmic discount.

In $DCG$, the relevance judgments ($v_i$) used to calculate gain values are commonly determined ad-hoc. The ad-hoc decision of relevance judgments may be problematic, as different judgments of relevance may derive different outcomes~\cite{Voorhees2001}. Therefore, the decision concerning this parameter is crucial for the evaluation process.

We find that previous work has neglected the impact of the numerical scale of ranked items' scores. In applications of $DCG$, relevance judgments are usually defined as integer numbers in an increasing manner depicting levels. This approach may discard important information concerning the degree of divergence between items' scores, potentially granting more or less relevance than in fact exists between ranked items, expressed by their respective true scores.

Therefore, in scenarios where the evaluation should also consider these score discrepancies between items in the ranking, we propose an automated, data-driven approach, for the definition of a degree of divergence relevance function, used to compute the relevance judgments ($v_i$) of each item in a given set, then incorporated in the metric $DCG$.

\section{Proposal}\label{sec:solution}

Let $X$ be a set of $n$ items, $Y$ the respective set of scores and $V$ a set of relevance judgments $v_1, \ldots, v_n \in V$ associated to $x_1, \ldots, x_n \in X$. Given a ranking function \textit{f} that orders $X$ according to the respective $Y$ scores in a decreasing fashion, the resulting ranking list $x_{(1)}^{f}, \ldots, x_{(n)}^{f}$ satisfies the condition $y_{(1)}^{f} \geq \ldots \geq y_{(n)}^{f}$.

Our hypothesis is that the ad-hoc definition of relevance judgments in $nDCG$ may under- or over- estimate the results of the metric due to the disregard of the distribution of the ranked items' scores.

We illustrate this scenario with a concrete example. Let $n=100$ and $Y$ a distribution of uniformly generated scores between $1$ and $1000$. For example purposes, we define in an ad-hoc manner that the judgments of relevance for an instance space of $\left \{ 0,1,2,3 \right \}$ are as such: items with the top-10 scores have a relevance of $3$, the remaining items in the top-25 scores a relevance of $2$, the remaining items in the top-50 scores a relevance of $1$, and the following items a relevance of $0$.

Let us assume that a recommendation system proposes a ranking where the top 10-$^{th}$ and top-$11^{th}$ items are permuted. Given the uniform distribution of the generated scores in this example, the divergence of score $y$ between these two items should be residual. However, when computing $nDCG$ using the previous ad-hoc parameters for relevance, we incur in a penalization that is disproportionate to the concrete score divergence between the items. This situation may also occur amongst items with the same relevance judgment, since items with a considerable score divergence may be attributed the same ad-hoc relevance. 

These examples coarsely show how the ad-hoc definition of relevance in $nDCG$ may incur in a disproportionate penalization of the evaluation when considering the degree of divergence amongst items' scores. Furthermore, they show the interest of using a principled approach to the definition of relevance judgments, motivating our proposal described in the next section.

\subsection{Degree of Divergence Relevance Function}\label{subsec:proposal}

The approach proposed in this paper consists of the interpolation of relevance judgments, based on the degree of divergence between ranked items' scores. The interpolated relevance judgments are used to calculate gain values used in the evaluation metric $nDCG$. In this paper we stipulate that relevance judgments are bounded by $\left [ 0,1 \right ], \mathbb{R}$, where $0$ is least relevant, and $1$ is most relevant.

To interpolate relevance judgments, we use Piecewise Cubic Hermite Interpolating Polynomials~\cite{Dougherty1989} (\textit{pchip}) in order to define the appropriate relevance function. This is done by interpolating a set of relevance judgments using specified control points. The choice of interpolation method was based on its simplicity and effectiveness in interpolating discrete data. Moreover, by restricting the first derivative values at the control points they are capable of preserving local positivity, monotonicity and convexity of the data. These are most convenient properties in the context of our target scenarios, as we want to induce a continuous function reproducing, as closely as possible, the control points provided.

We note that the control points are domain-dependent. This allows for our approach to meet the specifics of each domain. Nonetheless, for the purposes of this paper we use as control points the minimum, median and maximum value of $Y$ scores, introducing the following constraints to the interpolation process:

\begin{enumerate}
    \item All items $x_1,\ldots,x_n$ with a respective score $y_1,\ldots,y_n$ smaller than the median score have a relevance of $0$ (\textit{i.e.} non-relevant): the minimum and median score have a relevance judgment of $0$;
    \item For the remaining items, relevance is interpolated from $\left [ 0,1 \right ]$, where the item with median score $\tilde{Y}$ has a relevance of $0$ and the item with maximum score has a relevance of $1$;
\end{enumerate}

In case there are items with extremely large scores, this formulation may not be enough to correctly bias the evaluation of proposed rankings. In effect, in such situations we need to make sure these items with extremely high scores get properly ranked. Therefore, one wants to penalise more seriously any ranking errors involving these cases. With this goal in mind, we bias our interpolation function to give these extremes a very high relevance judgement, and to clearly distinguish their relevance from that of non-extreme items. The box plot rule~\cite{Cleveland1993} says that any value larger than $Q_3 + 1.5\times IQR$ (known as the upper whisker), where $Q_3$ is the third quartile of a distribution and $IQR= Q_3 - Q_1$ is the inter-quartile range, should be considered extreme, i.e. an outlier. 

We use this rule to determine if there are extreme values in our scores. If it is the case, we add a new control point to bias our interpolation function to assign relevance judgements that penalise ranking errors involving the items with these extreme scores. More specifically, the item with a score corresponding to the upper whisker will obtain a relevance judgement of $1 - \alpha$, where $\alpha$ is defined as

\begin{equation}
\alpha = \frac{max(Y) - (Q3(Y) + 1.5 IQR(Y))}{max(Y) - min(Y)}
\end{equation}

\begin{figure}[!h]
    \centering
    \includegraphics[width=\linewidth]{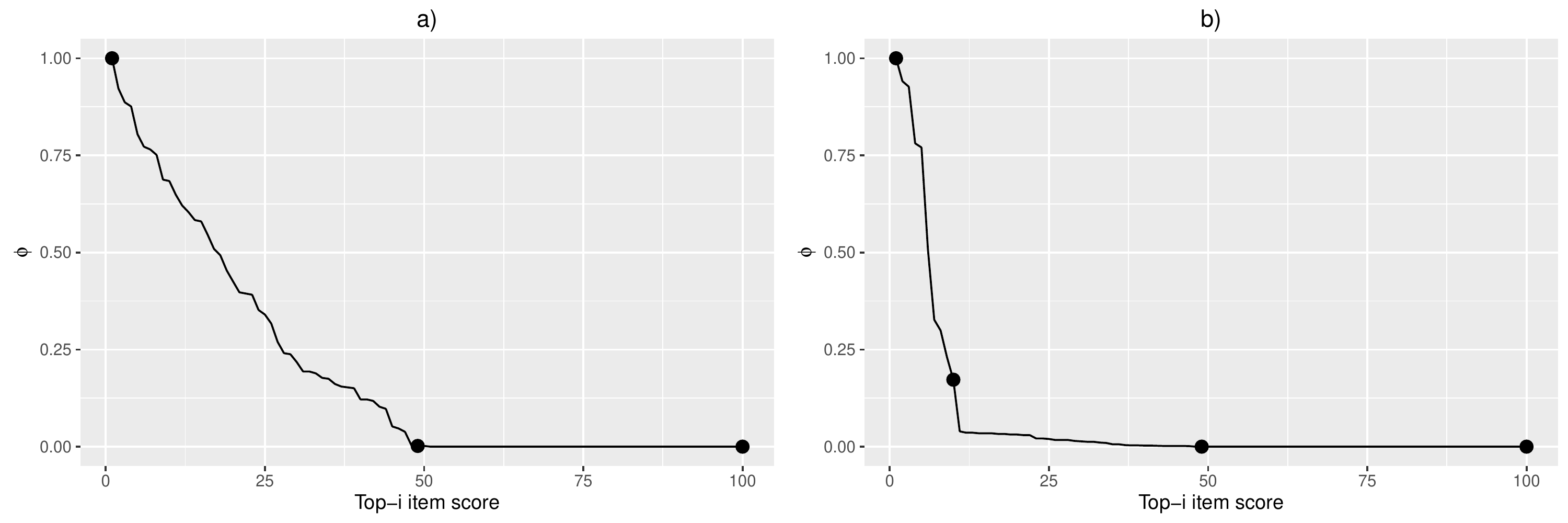}
    \caption{Examples of relevance functions generated by our proposal for cases without outliers (a) and for cases with outliers (b) using artificially generated data. The points in the examples represent the control points.}
    \label{fig:relexample}
\end{figure}

To illustrate, in Figure~\ref{fig:relexample} we show two examples of relevance functions using our approach using artificially generated data: one for cases without outliers, and another for cases with outliers.

In summary, we define the relevance judgements of items, used in calculating gain values of $nDCG$, through the application of $pchip$ using the above mentioned control points. We call the resulting relevance function $\phi$, which is then used in the formula of $nDCG$. We will refer to this proposed variant as $nDCG_{\phi}$.

\section{Experimental Evaluation}\label{sec:exps}

In this section we present two sets of results regarding synthetic and real-world application data. The synthetic data experiments show that in given situations of rank ordering, the impact of the degree of divergence in items' scores is observed, and the standard $nDCG$ may under- or over-estimate the results. By using real-world data, the objective is to show the difference between the standard $nDCG$ and $nDCG_{\phi}$, and the impact of our proposal. Both experiments use the cut-off version of $nDCG$ ($nDCG@k$). The cut-off version only considers the top $k$ items.

\subsection{Synthetic Data}

We generated rankings using the example described in Section~\ref{sec:solution} with the sets of items $X$ and score $Y$ of size $n=100$. The set of ad-hoc relevance judgments $V$ used by the standard $nDCG@k$ is also defined as in the example from Section~\ref{sec:solution}. In this experiment we set $k=10$ (\textit{i.e.} $nDCG@10$). 

Two types of distribution of the $Y$ scores are included: balanced and imbalanced. In the former, $Y$ scores are uniformly generated within the interval of $\left [ 1,1000 \right]$. As for the latter, 90\% of the scores are generated within the interval of $\left [ 1,100 \right ]$ and 10\% within the interval of $\left [ 100,1000 \right ]$.

We describe two scenarios that illustrate concrete situations where our stated hypothesis may be proven. Consider an ideal rank $R = \left ( r_1,r_2,\ldots,r_{10} \right )$ where $i$ is the position of a ranked item, $r_i$.  The testing scenarios are the following:

\begin{itemize}
    \item Scenario 1: to exemplify a permutation of items with different ad-hoc relevance judgments, $r_{10}$ is permuted with $r_{11}$: $R^{\prime}= \left ( r_1,r_2,\ldots,r_{9},r_{11} \right )$;
    \item Scenario 2: corresponds to an inverted ideal rank $R^{\prime\prime} = \left ( r_{10},r_9,...,r_1 \right )$.
\end{itemize}  

\begin{figure}[!h]
    \centering
    \includegraphics[width=8cm]{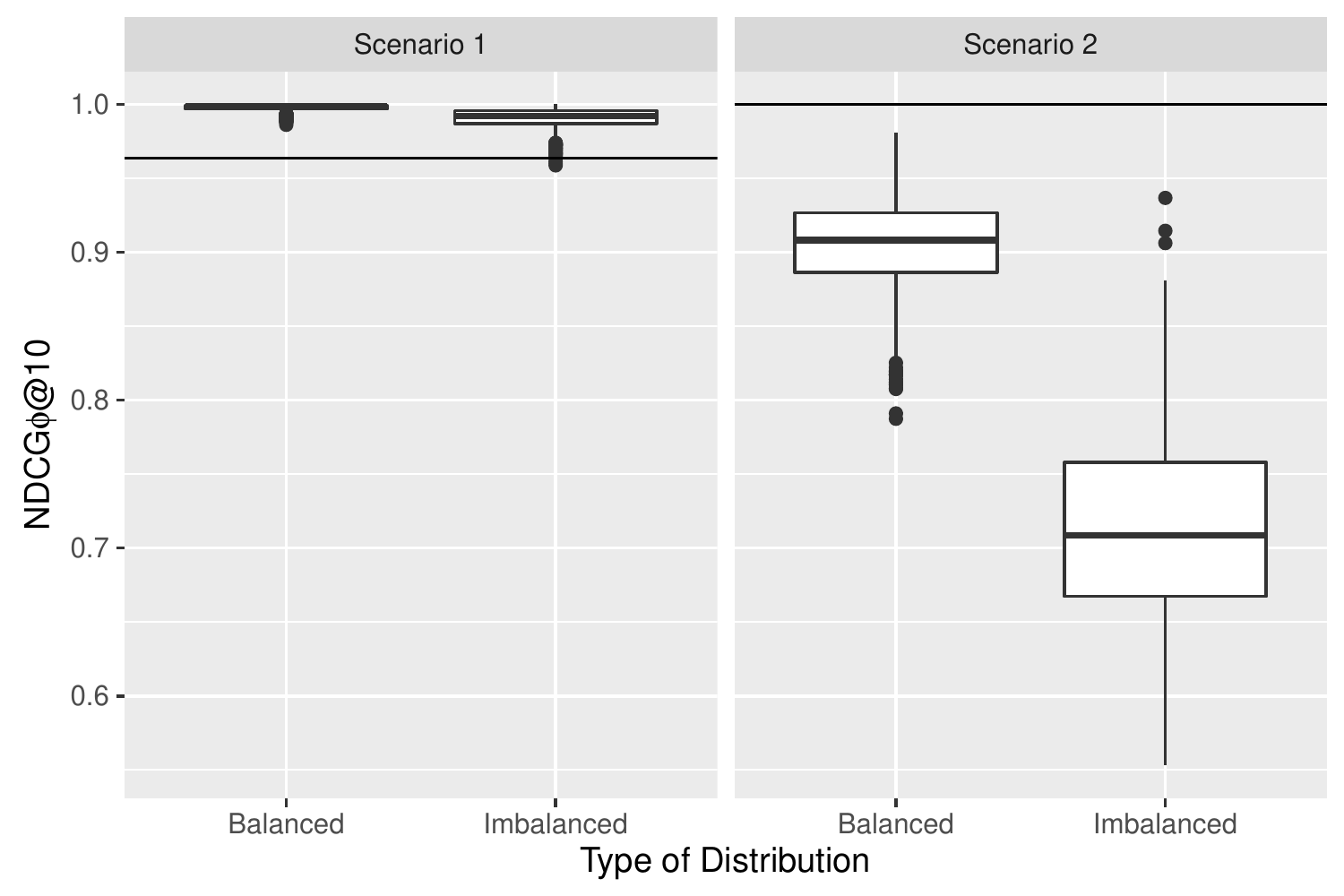}
    \caption{Comparison between $nDCG@10$ (horizontal line) and $nDCG_{\phi}@10$ (box plot) in two scenarios, by type of distribution.}
    \label{fig:synthetic}
\end{figure}

For each scenario and type of distribution we generated 1000 random samples of $Y$ scores. For each of these cases we calculate the respective $nDCG@10$ (horizontal line) and $nDCG_{\phi}@10$ (box plot) scores. Please note that as $nDCG@10$ is insensitive to the $Y$ scores and the relevance judgments are fixed (top-10 items' scores have a relevance judgment of $3$, remaining items in top-25 items' scores have a relevance judgment of $2$, $\ldots$), the evaluation score is constant for the 1000 samples. In Figure~\ref{fig:synthetic} we present the outcome of the compared evaluation between $nDCG@10$ and $nDCG_{\phi}@10$. 

In Scenario 1, where items in position $10$ and $11$ are permuted, we observe similar results by both metrics. Nevertheless, a fluctuation of $nDCG_{\phi}@10$ results shows that, in the majority of cases, the standard $nDCG@10$ over-estimates the impact of the ordering error in both the balanced and imbalanced distributions. In Scenario 2, depicting an inverted ideal rank, results show that $nDCG_{\phi}@10$ varies considerably. This shows the impact of the underlying distribution of ranked items' scores, when considered. According to the results in this scenario, the standard $nDCG@10$ metric under-estimates the impact of the ordering errors in both the balanced and imbalanced distributions. Moreover, we note that the impact of our proposal concerning the imbalanced distribution is much greater. This outcome shows the impact that the type of distribution has on both evaluations. As expected, when the scores are more imbalanced the ranking errors tend to have a greater impact on the evaluation, and the $nDCG_{\phi}@10$ metric is able to capture this in terms of overall evaluation of the rankings.

Given our hypothesis that the ad-hoc definition of relevance judgments in $nDCG@10$ may under- or over-estimate the results of the metric due to the disregard of the distribution of ranked items' scores, results from this experiment show that our hypothesis has empirical grounds.

\subsection{Real-World Data}\label{subsec:realworld}

In this set of experiments, we used data from Google News and Twitter. News items were retrieved from Google News over a timespan of six months (between June 1st and December 31st 2015) for four different topics: \emph{economy, microsoft, obama,} and \emph{palestine}. During the collection period, a query for each topic was posed every 30 minutes, and the top-100 news were retrieved. For each news item retrieved, we collected its position in the Google News ranking.

Using the Twitter API\footnote{Twitter API: \url{https://dev.twitter.com/docs/api}. The \textit{count} method was deprecated on 20$^{th}$ of November, 2015.} we retrieved the number of times the news was tweeted in the two days following its publication. The two days' limit was decided considering the work of Yang and Leskovec~\cite{Yang2011}. The number of times a news item was tweeted is used as its score, and considered a proxy for its popularity for end-users. For each Google News ranking, the ground-truth is given by a descending order ranking of the number of tweets obtained by each ranked item. This data set contains approximately 9.795 rankings per topic and a total of 107.590 news items where 6.468 items (6\%) were not published in Twitter. 

\begin{figure*}[!h]
    \centering
    \includegraphics[width=\linewidth]{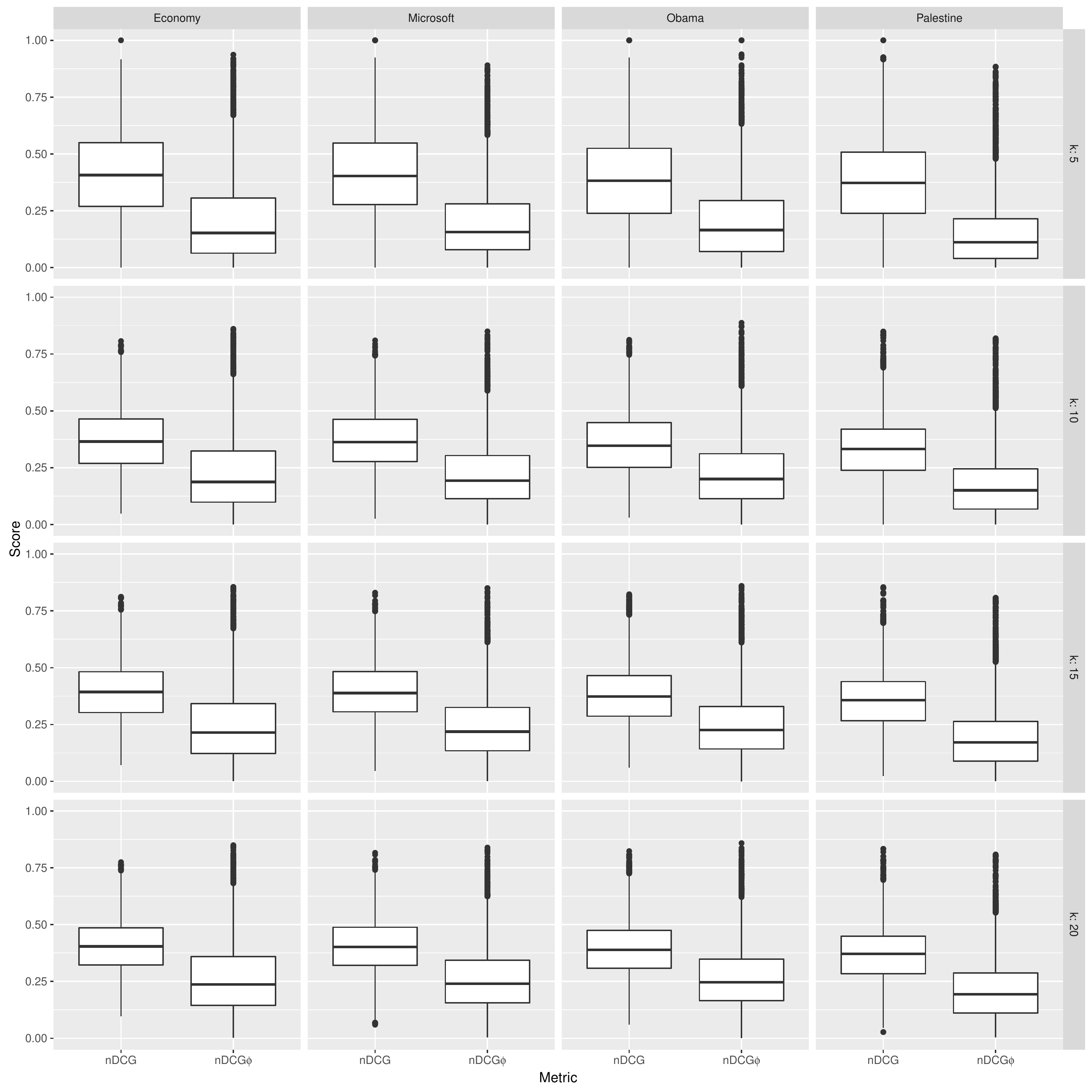}
    \caption{Results concerning the metrics $nDCG$ and $nDCG_{\phi}$ for several values of $k$ in all four topics of the real-world data.}
    \label{fig:realworldexp}
\end{figure*}

An evaluation of each ranking produced by Google News using $nDCG@k$ and $nDCG_{\phi}@k$ is depicted in Figure~\ref{fig:realworldexp}. The objective is to illustrate the overall differences between both metrics for different values of $k$ in all four topics: $5$, $10$, $15$ and $20$. The ad-hoc relevance judgments used in the standard $nDCG$ are those described in Section~\ref{sec:solution}.

Results show that the average evaluation of the rankings using the standard ad-hoc $nDCG$ metric consistently shows better evaluation scores in comparison to $nDCG_{\phi}$. Considering that the $nDCG_{\phi}$ metric accounts for the items' scores divergence, this analysis shows that despite the increasing value of $k$ and the topic evaluated, the $nDCG$ metric tends to under-estimate the impact of ordering errors in the rankings. 

Therefore, the results obtained show that our approach is useful for evaluation scenarios where items' scores are available and the divergence between scores is an important a factor. The use of $nDCG_{\phi}$ allows for an evaluation that shows a greater sensibility to the degree of divergence between ranked items, in comparison to the ad-hoc definition of relevance in the standard $nDCG$.

\section{Discussion}\label{sec:discussion}

Considering the results of the experimental evaluation in both the synthetic and real-world settings, we have shown the interest of our proposal. By proposing a data-driven approach for the definition of relevance functions and its application in the popular $nDCG$ evaluation metric we allow for a much finer-grain process for evaluating rankings. Also, we show that our proposal is capable of correcting to some extent the issues of under- and over- evaluation related to the common use of ad-hoc relevance judgments in the $nDCG$ metric.

\begin{figure*}[!h]
    \centering
    \includegraphics[width=\linewidth]{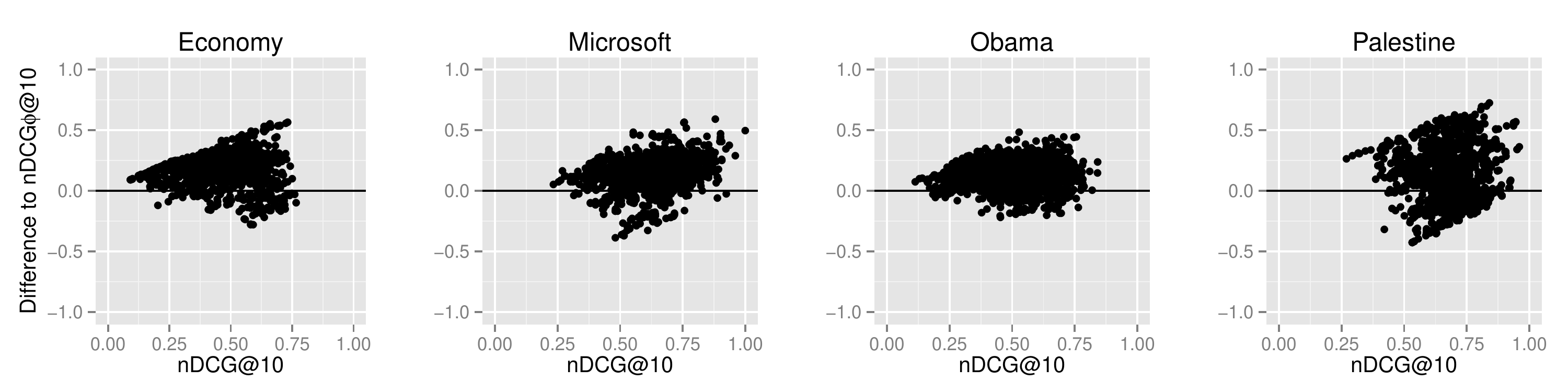}
    \caption{Difference between metrics ($nDCG@10-nDCG_{\phi}@10$) on rankings of news retrieved from Google News.}
    \label{fig:realworld}
\end{figure*}

Consider the analysis shown in Figure~\ref{fig:realworld}. In this Figure we use a sample (2000 rankings) of the results from the evaluation presented in Section~\ref{subsec:realworld} and compare the scores of both $nDCG@10$ and $nDCG_{\phi}@10$ metrics, in all topics. It illustrates the score differences of both metrics (the $y-$axis shows the value of $nDCG@10-nDCG_{\phi}@10$).

This shows that both metrics tend to evaluate the recommended rankings differently. Analyzing the differences between both metrics, we learn that $nDCG@10$ tends to assign a higher evaluation score to rankings, depicted by the differences between the two metrics being more frequently positive. Also, it shows that this difference tends to increase with higher evaluation scores from $nDCG@10$. The fact that we observe these differences mean that in the real world scenario used, the variability of the scores that originate the rankings is present and significant, otherwise our proposal would not differ so much from standard $nDCG@10$. Given the increasing throughput of online data and the changes (\textit{i.e.} seasonality, unexpected behaviour) in user behaviour, this also shows the adaptability of our proposal and its contribution to the problem of evaluating rankings.

Finally, we should clarify that in light of the scenarios described in this paper, the ad-hoc definition of relevance judgments commonly used with the $nDCG$ metric is naturally prone to the under- or over-evaluation issues described. This is due to the level-like depiction of relevance commonly used. Nonetheless, the interest of our proposal lies on it being a principled way of depicting relevance judgments and thus bypassing the issues raised by the common ad-hoc definition.

\section{Conclusions}\label{sec:conclusions}

In this paper we study the performance of the metric Normalized Discounted Cumulative Gain ($nDCG$) in light of the numerical scale and distribution of ranked items' scores. We show that the standard $nDCG$ presents evidence of under- and over-estimating ordering errors in rankings due to the use of ad-hoc relevance judgements and the disregard for the underlying distribution of the items' scores, when available and considered. We propose a data-driven approach capable of, under different types of distributions, interpolating the relevance of the items in a ranking, based on the degree of the divergence amongst the items' scores. Two sets of experiments were carried out, empirically showing the interest of our proposal resorting to specific scenarios, and by demonstrating its usefulness in a real-world scenario, in comparison to the ad-hoc definition of relevance judgments used in standard $nDCG$ metric. Future work will involve the extension of experiments to enable a more detailed explanation for the difference in results between the two metrics. Additionally, we wish to study the possibility of using this approach with other ranking evaluation metrics.

For reproducibility, all code (written in \textbf{R}) and data necessary to replicate the results are available in the Web page \url{http://tinyurl.com/h454dry}.

\section{Acknowledgments}
This work is financed by the ERDF – European Regional Development Fund through the COMPETE 2020 Programme within project POCI-01-0145-FEDER-006961, and by National Funds through the FCT – Fundação para a Ciência e a Tecnologia (Portuguese Foundation for Science and Technology) as part of project  UID/EEA/50014/2013. It is also financed by the Project "Tec4Growth - Pervasive Intelligence, Enhancers and Proofs of Concept  with  Industrial Impact/NORTE-01-0145-FEDER-000020", which is financed by the  North  Portugal  Regional  Operational  Programme  (NORTE  2020), under the PORTUGAL 2020 Partnership Agreement, and through the European Regional Development Fund (ERDF). 
The work of N. Moniz is supported by a PhD scholarship of FCT (SFRH/BD/90180/2012).

\bibliographystyle{splncs03}

\end{document}